\documentclass{icrc2009}

\usepackage{graphicx}   % for including figures
\usepackage{fixltx2e}
\usepackage{url}

\newcommand{\shorttitle}[1]%
{\markboth{Proceedings of the 31\MakeLowercase{$^{st}$} ICRC, {\L}\'{o}d\'{z} 2009}{#1} }
\newcommand{\etal}{\MakeLowercase{\textit{et al. }}} % "et al."

\begin{document}

\title{Radio emission of extensive air shower at CODALEMA: Polarization of the radio emission along the $\mathbf{v}\times\mathbf{B}$ vector}
\author{\IEEEauthorblockN{Colas Rivi\`ere\IEEEauthorrefmark{1} for the CODALEMA Collaboration}
\IEEEauthorblockA{
\IEEEauthorrefmark{1}Laboratoire de Physique Subatomique et de Cosmologie, UJF, INPG, CNRS/IN2P3, Grenoble, France
}
}

\shorttitle{C. Rivi\`ere \etal Radio emission polarization with CODALEMA}
\maketitle

\begin{abstract}
Cosmic rays extensive air showers (EAS) are associated with transient radio emission, which could provide an efficient new detection method of high energy cosmic rays, combining a calorimetric measurement with a high duty cycle. The CODALEMA experiment, installed at the Radio Observatory in Nan\c{c}ay, France, is investigating this phenomenon in the $10^{17}$~eV region. One challenging point is the understanding of the radio emission mechanism. A first observation indicating a linear relation between the electric field produced and the cross product of the shower axis with the geomagnetic field direction has been presented (B. Revenu, this conference). We will present here other strong evidences for this linear relationship, and some hints on its physical origin.
\end{abstract}

\begin{IEEEkeywords}
 CODALEMA, radio detection, polarization
\end{IEEEkeywords}

\section{Introduction}
Radio detection is a promising detection technique of extensive air shower as it could provide at low-cost a calorimetric measurement of the shower with a high duty cycle. The first exploration phase of this technique is described in Allan's review of 1971~\cite{allan}. The observations of the electric field in the East-West polarization are there summarized with the following equation of the electric field produced on the ground by an EAS:
\begin{eqnarray}
  \mathcal{E}_{\nu}&{}={}&20 \left(\frac{E_p}{10^{17}\textnormal{~eV}}\right) \sin\alpha \cos\theta \nonumber\\
  &&{\times}\:\exp\left( \frac{-R}{R_0\left(\nu,\theta\right)} \right) 
  \qquad \frac{\mu\textnormal{V}}{\textnormal{m MHz}}
  \label{allan_formula}
\end{eqnarray}
$E_p$ stands for the primary cosmic ray energy in eV, $\alpha$ is the angle between the shower axis and the geomagnetic axis, $\theta$ the zenith angle and $R$ radial distance to the shower axis. The value $R_0$ is estimated for $\theta<35°$ and at two frequencies: 100$\pm$10~m at 55~MHz and $\sim$140~m at 32~MHz.

The radio detection technique has been neglected for a while in favor of the fluorescence and surface detectors developments. Nowadays, progress of fast and low cost electronics offers a new insight for radio detection. The experiments CODALEMA in France and LOPES in Germany are currently dedicated to the study of this technique in the $10^{17}$~eV region~\cite{CODALEMA,LOPES}, along with other preliminary experiments in Argentina~\cite{ARENA_Benoit,ARENA_Jose}. This experimental work is supported by a renewal in theories, with several emission models and implementations such as the geosynchrotron emission, the boosted Coulomb field or the transverse current~\cite{falcke2003dre,huege,gousset,lecacheux,scholten}.

%%%%%%%%%%%%%%%%%%%%%%%%%%%%%%%%%%%%%%%%%%%%%%%%%%%%%%%%%%%%%%%%%%%%%%%%%%
%%%%%%%%%%%%%%%%%%%%%%%%%%%%%%%%%%%%%%%%%%%%%%%%%%%%%%%%%%%%%%%%%%%%%%%%%%
\section{The $\mathbf{v}\times\mathbf{B}$ polarization vector}
The synchrotron emission of an electron moving through the Earth's magnetic field can be calculated using the general formula of an accelerated relativistic particle, with $\mathbf{v}$ representing here the velocity of the particle and $\mathbf{n}$ the direction of the observer:
  \begin{eqnarray}
    \mathbf{E}&{}={}&\frac{e}{4\pi\epsilon_0}
    				\left[\frac{\mathbf{n}-\mathbf{v}}
    					{\gamma^2(1-\mathbf{v}\cdot\mathbf{n})^3R^2}\right]_{ret} \nonumber\\
    			&&{+}\:\frac{e}{4\pi\epsilon_0c}
				\left[\frac{\mathbf{n}\times\{(\mathbf{n}-\mathbf{v})\times\dot{\mathbf{v}}\}}
					{(1-\mathbf{v}\cdot\mathbf{n})^3R}\right]_{ret}
    \label{jackson_formula}
  \end{eqnarray}
The second term of the equation $\mathbf{E_2}$ represents the geosynchrotron emission when $\dot{\mathbf{v}}=\frac{e}{\gamma m}\mathbf{v}\times\mathbf{B}$ is the Lorentz acceleration in the magnetic field $\mathbf{B}$. In the special case of an observation point lying on the particle motion axis, we have $\mathbf{v}\parallel\mathbf{n}$:
  \begin{eqnarray}
    \mathbf{E_2}
	&{}\propto{}&
	    \mathbf{n}\times\{(\mathbf{n}-\mathbf{v})\times(\mathbf{v}\times\mathbf{B})\} \nonumber\\
    &{}\propto{}&
		\underbrace{\{\mathbf{n}.(\mathbf{v}\times\mathbf{B})\}}_{=0\textnormal{~on the axis}}(\mathbf{n}-\mathbf{v})
		-\{\mathbf{n}.(\mathbf{n}-\mathbf{v})\}(\mathbf{v}\times\mathbf{B}) \nonumber\\
    &{}\propto{}&
		-(1-v)(\mathbf{v}\times\mathbf{B}) \nonumber\\
    &{}\propto{}&
	    -\mathbf{v}\times\mathbf{B}
    \label{vB_formula}
  \end{eqnarray}
thus the synchrotron electric field produced near the particle motion axis is at first order proportional to this cross product.

The transposition can be made to an EAS, where $\mathbf{v}$ represents from now on the direction of the shower axis, and $\mathbf{B}$ corresponds to the geomagnetic field. The $\mathbf{E}\propto-\mathbf{v}\times\mathbf{B}$ relation is conserved in the case of a realistic EAS, as checked with a full Monte Carlo simulation of the geosynchrotron emission in a modified version of the program AIRES~\cite{AIRES,ReAIRES}. These simulation results are presented in another contribution to this conference (C. Rivi\`ere~\cite{ICRC_Colas_aires}). Other emission models may give the same dependence.

We can note that the modulus of the $\mathbf{v}\times\mathbf{B}$ vector is $\sin\alpha$, as in Eq.~\ref{allan_formula}. However, the $\sin\alpha$ corresponds here to the total field amplitude, not to the East-West polarization. The East-West component of the fields is then given by the projection of the $\mathbf{v}\times\mathbf{B}$ vector on the EW axis.

%%%%%%%%%%%%%%%%%%%%%%%%%%%%%%%%%%%%%%%%%%%%%%%%%%%%%%%%%%%%%%%%%%%%%%%%%%
%%%%%%%%%%%%%%%%%%%%%%%%%%%%%%%%%%%%%%%%%%%%%%%%%%%%%%%%%%%%%%%%%%%%%%%%%%
\section{The CODALEMA experiment}
\begin{figure}[htbp]
  \centering
  \includegraphics[width=\columnwidth]{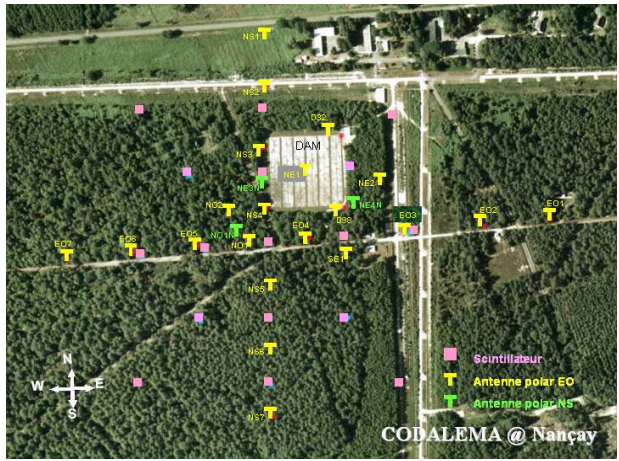}
%  \includegraphics[width=\columnwidth]{./img/nancayperiodeVI_BW}
% g b d h
  \caption{Layout of the CODALEMA experimental setup. Scintillator stations are represented in pink squares, EW antennas in yellow, NS antennas in green.}
  \label{CODALEMA_overview}
\end{figure}

The CODALEMA experimental setup, located in Nan\c{c}ay, France, is currently composed of an array of 17 particle detectors overlapped with an array of 24 wide band active dipoles: 21 oriented along the East-West (EW) direction and 3 oriented along the North-South (NS) direction, disposed as shown on Fig.~\ref{CODALEMA_overview}.
The whole acquisition is triggered by the particle detector array, which also provides reference information on the shower characteristics. After offline filtering of the radio signals in the 23--83~MHz band, the radio information is reconstructed and then confronted to the surface detector data.
Additional information about the CODALEMA setup and data analysis can be found in Ref.~\cite{AP2009}.

A first observation indicating a linear relation between the electric field produced and the cross product $\mathbf{v}\times\mathbf{B}$ of the shower axis $\mathbf{v}$ with the geomagnetic field direction $\mathbf{B}$ is presented in another contribution to this conference~\cite{ICRC_Benoit}. We will detail bellow evidences of this relation. The following analysis is based on data taken during 453 effective days of stable acquisition with this configuration of 24 antennas. A radio event is an event with at least three fired antennas in order to reconstruct a radio shower plane. A radio detected event is a radio event which matches the coincidence criteria with the ground detector (arrival direction difference $<20^\circ$ and time difference $<200$~ns).

%%%%%%%%%%%%%%%%%%%%%%%%%%%%%%%%%%%%%%%%%%%%%%%%%%%%%%%%%%%%%%%%%%%%%%%%%%
%%%%%%%%%%%%%%%%%%%%%%%%%%%%%%%%%%%%%%%%%%%%%%%%%%%%%%%%%%%%%%%%%%%%%%%%%%
\section{Experimental sky coverage}
\begin{figure}[htbp]
  \centering
  \includegraphics[width=0.49\columnwidth,viewport=15 17 260 245,clip]{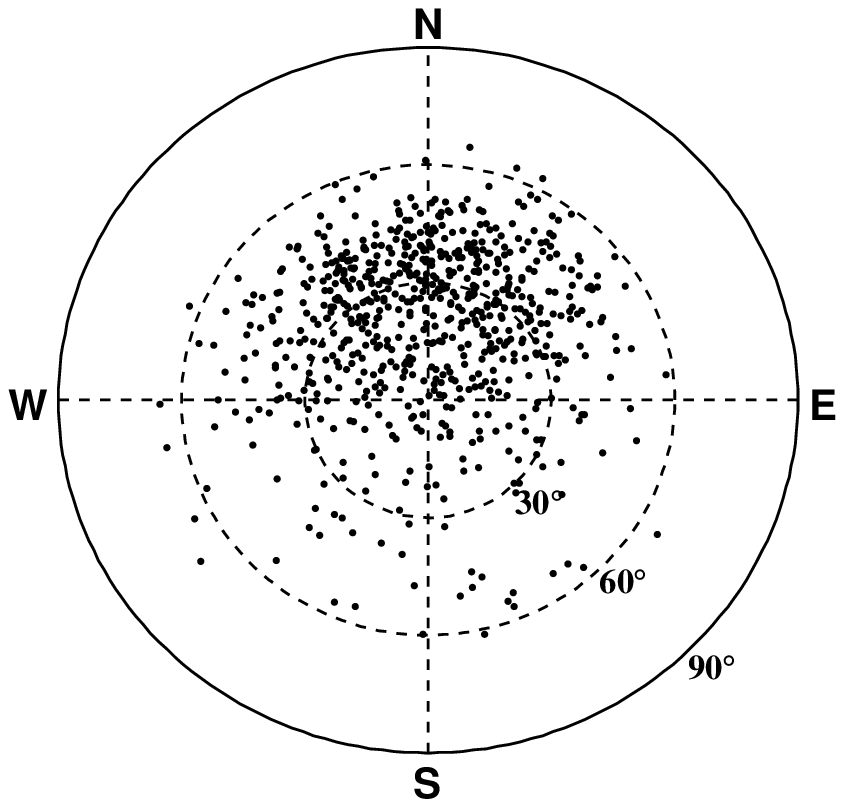}
  \includegraphics[width=0.49\columnwidth,viewport=15 17 260 245,clip]{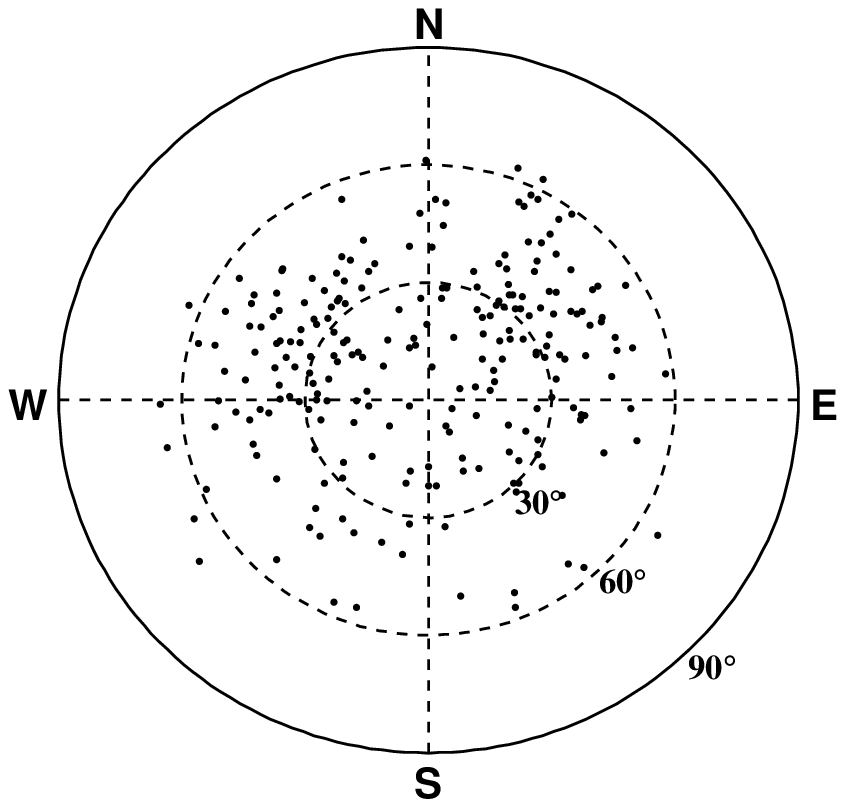}
%  \includegraphics[width=0.49\columnwidth,viewport=15 17 260 245,clip]{./img/sky_ew_BW}
%  \includegraphics[width=0.49\columnwidth,viewport=15 17 260 245,clip]{./img/sky_ns_BW}
% g b d h
  \caption{Arrival directions of the radio detected events containing at least one EW (left) or NS (right) polarized flagged signal.
    }
  \label{anisotropy}
  \centering
  \includegraphics[width=0.24\textwidth,viewport=13 17 272 256,clip]{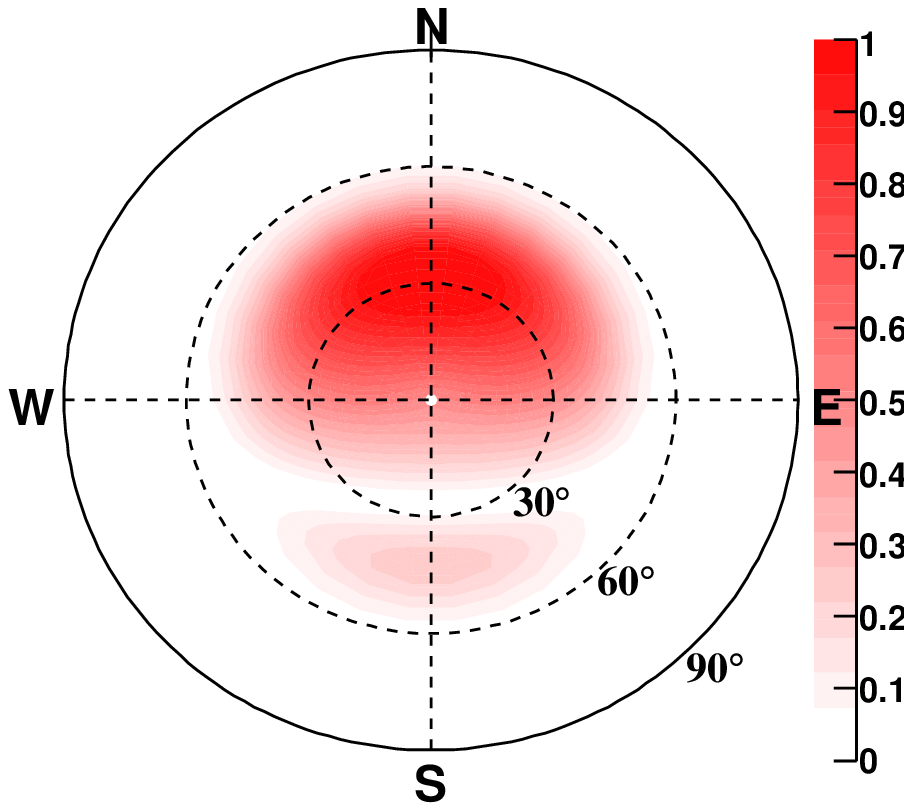}
  \includegraphics[width=0.24\textwidth,viewport=13 17 272 256,clip]{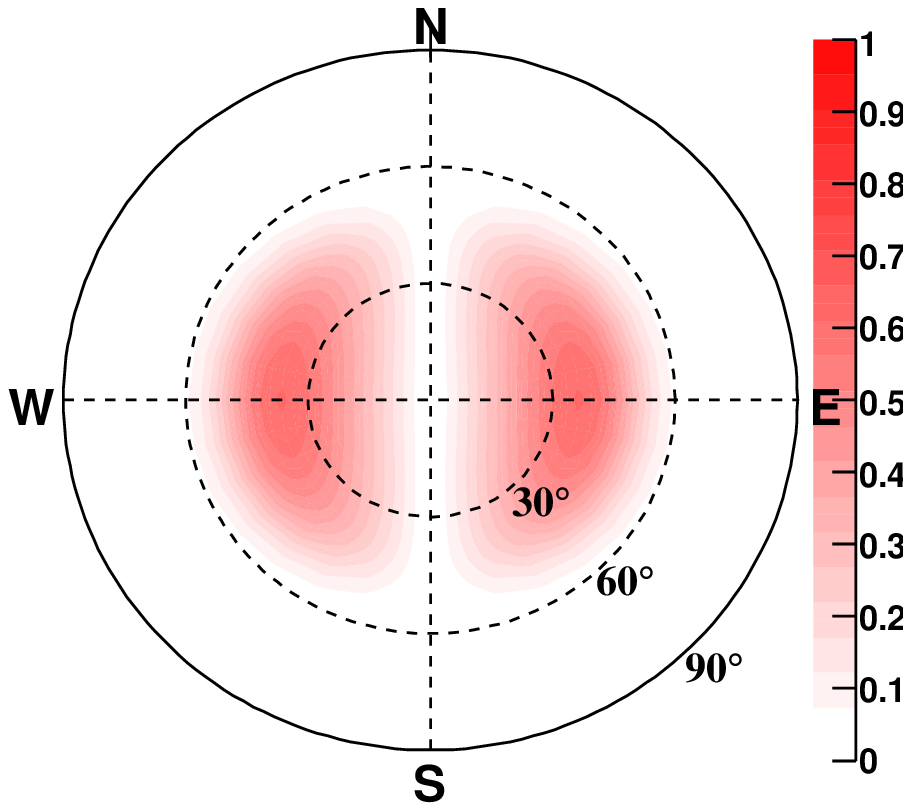}
%  \includegraphics[width=0.24\textwidth,viewport=13 17 272 256,clip]{./img/formula_ew_fabs_BW}
%  \includegraphics[width=0.24\textwidth,viewport=13 17 272 256,clip]{./img/formula_ns_fabs_BW}
% g b d h
  \caption{Product of CODALEMA particle detector acceptance above $10^{17}$~eV times the EW (left) or NS (right) component of the vector cross product (absolute values).
    }
  \label{accept_vB}
 \end{figure}

The arrival directions of the radio detected events present a strong anisotropy. The distributions of arrival directions are different when considering the EW or NS polarization, as shown on Fig.~\ref{anisotropy}. EW oriented antennas tends to see more showers coming from the North, whereas the NS oriented antennas tends to see showers coming from East and West.

The radio detection threshold of CODALEMA is around $10^{17}$~eV at 50\%. Let us consider the distribution of events above this energy\footnote{
$\frac{\textnormal{d}N}{\textnormal{d}\theta}=(a+b\theta)\cos\theta\sin\theta/(1+\exp((\theta-\theta_0)/\theta_1))$, with $a=44.96$, $b=0.57$, $\theta_0=49.18^\circ$ and $\theta_1=5.14^\circ$, cf.~\cite{ICRC_Benoit}}. If we multiply this distribution by the different projections of the cross product $\mathbf{v}\times\mathbf{B}$, we obtain the function plotted on Fig.~\ref{accept_vB}. Making two simple assumptions, this can be expected to represent an event rate distribution: i) the electric field is proportional to the vector cross product, ii) the probability of detection is proportional to the electric field. This is realistic here because we are close to the detection threshold.

We see that this simple model reproduces well the characteristics of the direction of the radio signals seen, in the two measured polarizations.

\begin{figure}[htbp]
  \centering
  \includegraphics[width=0.22\textwidth,viewport=30 20 800 800,clip]{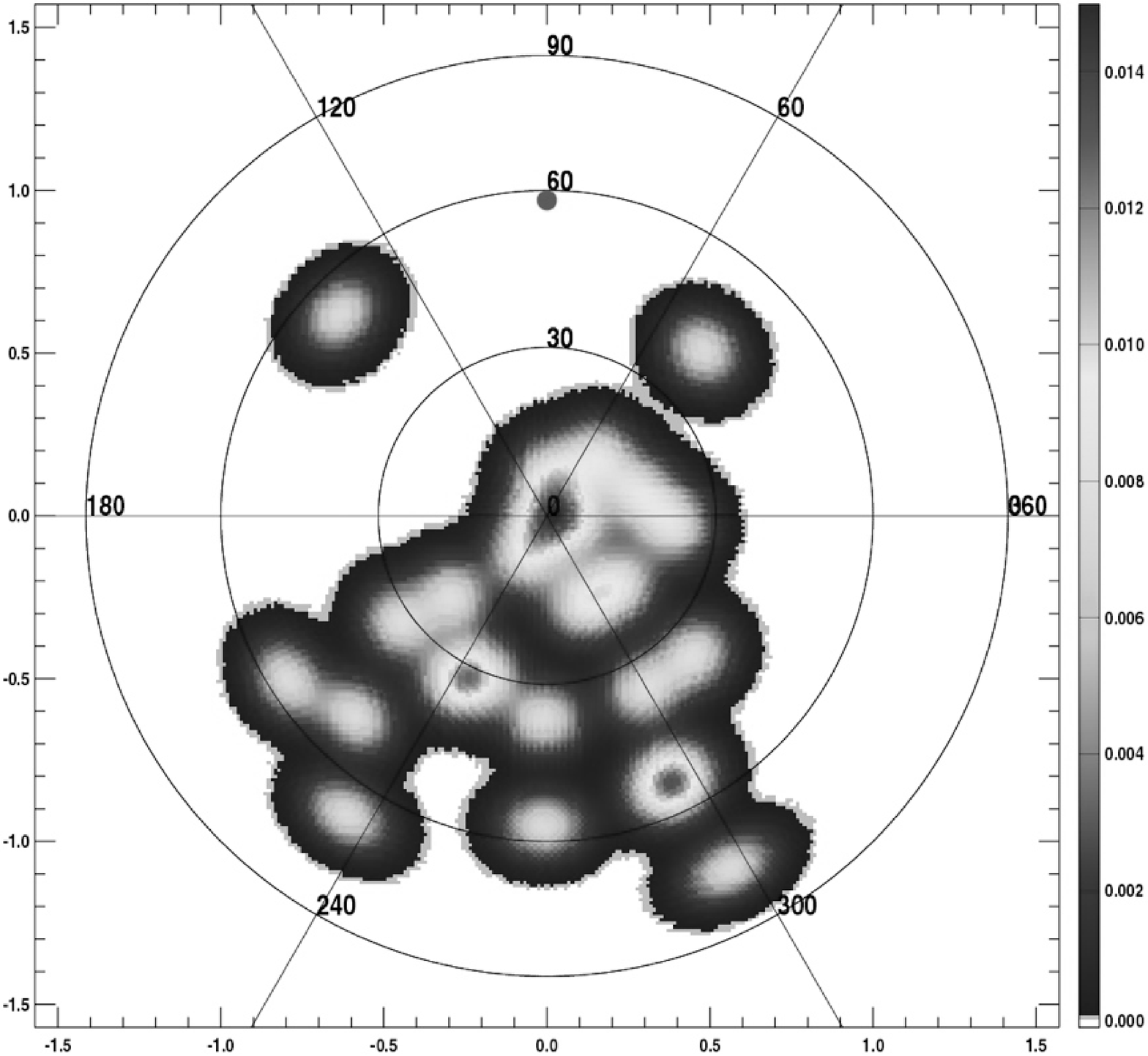}
  \includegraphics[width=0.25\textwidth,viewport=13 17 272 256,clip]{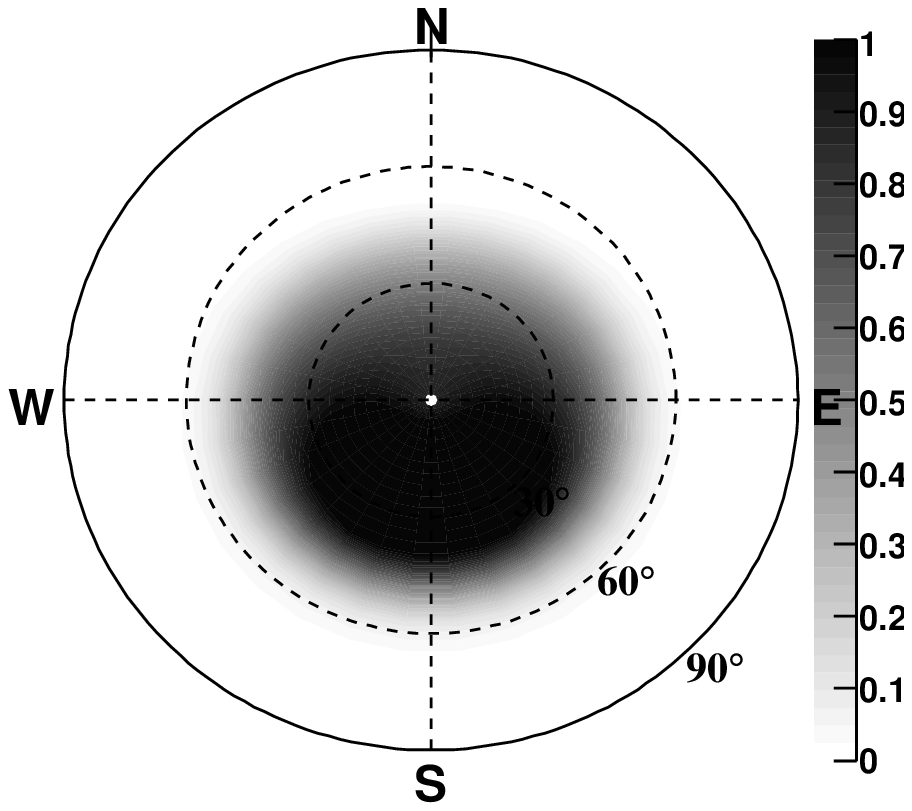} % Garder celle la, meme en couleur
% g b d h
  \caption{Left: arrival directions of the self triggered radio events detected in Auger (left). Right: Auger surface detector acceptance estimation multiplied by $|(\mathbf{v}\times\mathbf{B}/vB)_{EW}|$}
  \label{RAuger}
 \end{figure}

Similar results are found in Argentina with a different geomagnetic field orientation, $35^\circ$ to the North instead of $63^\circ$ to the South in Nan\c{c}ay. On Fig.~\ref{RAuger}, left, are shown the arrival direction of the events detected with autonomous radio detectors~\cite{ARENA_Benoit}. These events have been self triggered with the radio signals of the EW polarized antennas, and are in coincidence with events seen with the surface detector of the Pierre Auger Observatory. On the right side of Fig.~\ref{RAuger} is plotted the absolute value of EW component of the vector cross product multiplied by an estimate of the Auger zenith acceptance distribution. The depletion of events on the North is also explained with the same model.

%%%%%%%%%%%%%%%%%%%%%%%%%%%%%%%%%%%%%%%%%%%%%%%%%%%%%%%%%%%%%%%%%%%%%%%%%%
%%%%%%%%%%%%%%%%%%%%%%%%%%%%%%%%%%%%%%%%%%%%%%%%%%%%%%%%%%%%%%%%%%%%%%%%%%
\section{Detection efficiency}
\begin{figure}[htbp]
  \centering
  \includegraphics[width=0.8\columnwidth,viewport= 0 5 250 240,clip]{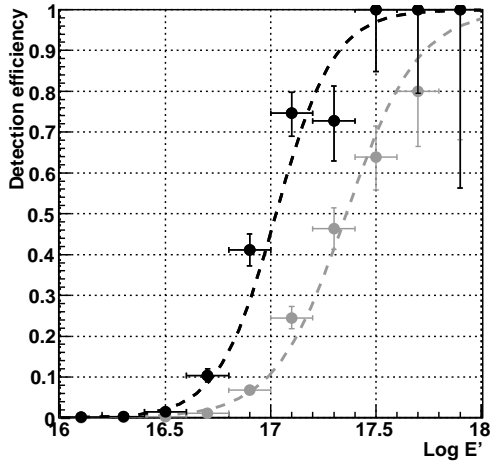}
% g b d h
  \caption{Radio detection efficiency, as a function of the energy E (gray) or of the corrected energy E' (black).} % Garder celle la, meme en couleur
  \label{efficiency}
\end{figure}

We consider here the events reconstructed with the array composed of the EW oriented antennas. The radio detection efficiency of the CODALEMA experiment is shown on Fig.~\ref{efficiency}. In gray, the detection efficiency is plotted against the energy estimated by the particle detector array, E. In black, the efficiency is plotted against $E'=E.|(\mathbf{v}\times\mathbf{B}/vB)_{EW}|$, the energy multiplied by the EW component of the vector cross product,  which represents an estimation of the electric field amplitude. As expected, the efficiency rises faster in the second case\footnote{Width of a Fermi-Dirac fit of the efficiency curve: 0.171 (E) and 0.137 (E'). Rq: 0.171 with a $\sin\alpha$ correction, and 0.157 with a $(1-\cos\alpha)$ correction.}. 

\begin{figure}[htbp]
  \centering
  \includegraphics[width=0.8\columnwidth,viewport= 0 5 250 240,clip]{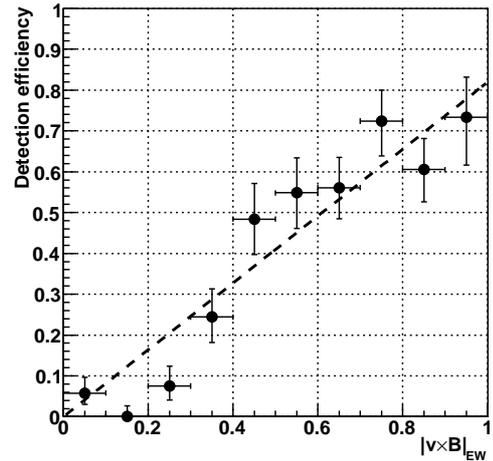}
% g b d h
  \caption{Radio detection efficiency vs. the EW component of the vector cross product $|(\mathbf{v}\times\mathbf{B}/vB)_{EW}|$, for the events above $10^{17}$~eV.}
  \label{efficiency_vB}
\end{figure}

Regardless the energy, the radio detection efficiency also increases with $|(\mathbf{v}\times\mathbf{B}/vB)_{EW}|$, as shown on Fig.~\ref{efficiency_vB} for the events above $10^{17}$~eV. The quasi linear tendency is in favor of the assumption made earlier that the detection efficiency is proportional to the electric field close to the energy threshold\footnote{The correlation is much weaker with $\sin\alpha$ or $(1-\cos\alpha)$.}.

%%%%%%%%%%%%%%%%%%%%%%%%%%%%%%%%%%%%%%%%%%%%%%%%%%%%%%%%%%%%%%%%%%%%%%%%%%
%%%%%%%%%%%%%%%%%%%%%%%%%%%%%%%%%%%%%%%%%%%%%%%%%%%%%%%%%%%%%%%%%%%%%%%%%%
\section{Signal sign}
\begin{figure}[htbp]
  \centering
  \includegraphics[width=0.24\textwidth,viewport=13 17 272 256,clip]{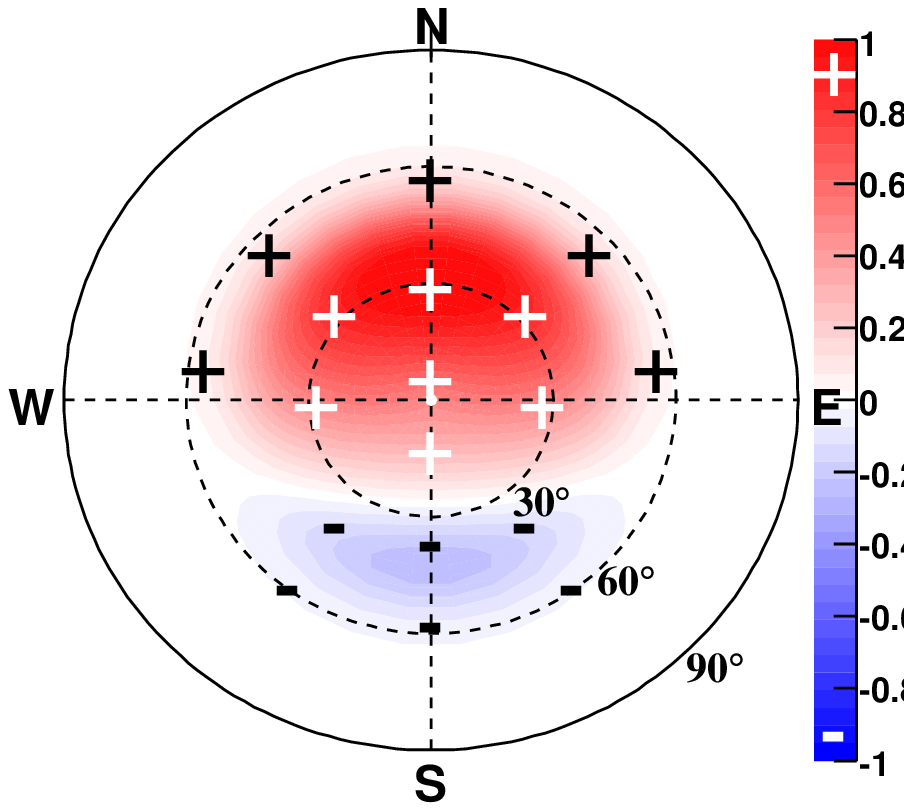}
  \includegraphics[width=0.24\textwidth,viewport=13 17 272 256,clip]{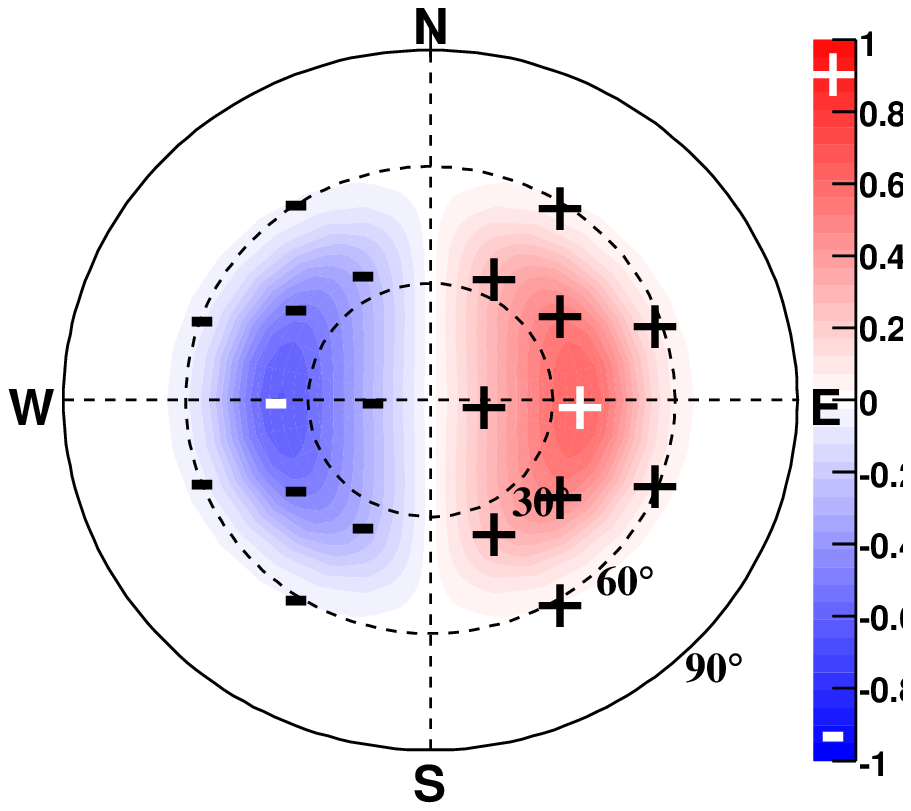}
%  \includegraphics[width=0.24\textwidth,viewport=13 17 272 256,clip]{./img/formula_ew_BW}
%  \includegraphics[width=0.24\textwidth,viewport=13 17 272 256,clip]{./img/formula_ns_BW}
% g b d h
  \caption{Sign of the EW (left) and NS (right) components of the vector cross product $-\mathbf{v}\times\mathbf{B}$. The scale is does not have importance here.}
  \label{signe_vB}
\end{figure}

The sign of the components cross product $-\mathbf{v}\times\mathbf{B}$ varies with the EAS arrival direction. The dependences of the EW and NS components are shown on Fig.~\ref{signe_vB}. The EW (resp. NS) component is positive on North (East) and negative on South (West).
\begin{figure*}[htbp]
  \centering
  \includegraphics[width=0.49\textwidth]{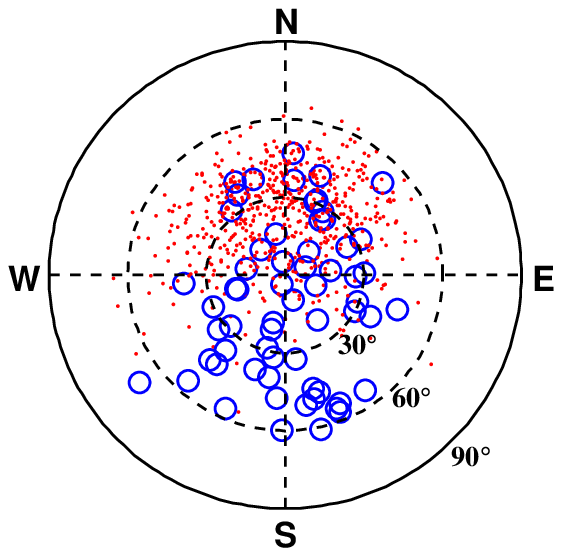}
  \includegraphics[width=0.49\textwidth]{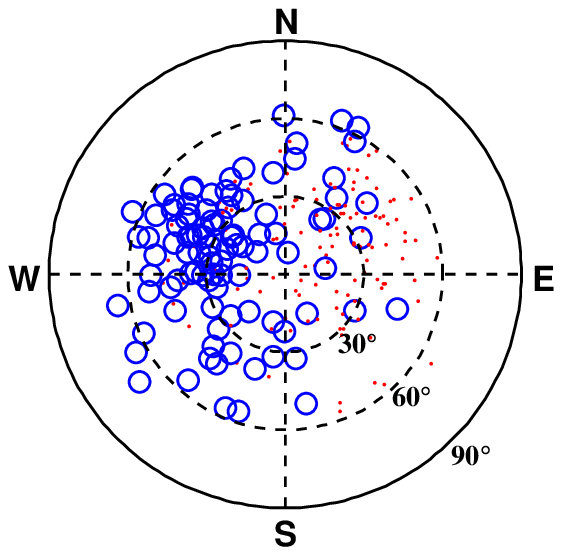}
%  \includegraphics[width=0.49\textwidth]{./img/signe_exp_ew_BW}
%  \includegraphics[width=0.49\textwidth]{./img/signe_exp_ns_BW}
% g b d h
  \caption{Experimental signs of the radio signals seen in Nan\c{c}ay, with the EW (left) or NS (right) oriented antennas. Red dots represent positive signals and blue circles negative ones.}
  \label{signe_exp}
\end{figure*}

Experimentally, one antenna signal is defined as the extremum of a time signal filtered in the 23--83~MHz band\footnote{Filtering of simulated signals indicates that the sign is conserved during the process, but anyway the sign flip with the arrival direction is more important than the absolute sign.}. The sign of the signal is simply the sign of this extremum. For an event with several antennas fired for the considered polarization, the sign of the event is chosen to the majority sign among the different signals. This reduces the influence of noise, and is of course more efficient in EW polarization than in NS as more antennas are available. The EW component of $-\mathbf{v}\times\mathbf{B}$ is also generally bigger than its NS component, thus the measurement of the sign is generally easier in the EW polarization.

The experimental signs of the radio detected events are shown on Fig.~\ref{signe_exp}. In the EW polarization, events coming from North generally have a positive sign and events coming have a negative sign. The apparently uniform distribution of the negative event is simply a statistic effect where appear the negative events coming from South plus a small fraction of the very numerous events coming from North. In the NS polarization, events coming from East are generally positive and events coming from West are generally negative. The overall distributions of the sign with the arrival direction is the same as the sign distribution of $-\mathbf{v}\times\mathbf{B}$: the agreement is 93\% for the EW polarization (with 19 antennas maximum), and 78\% for the NS polarization (with 3 antennas maximum).

%%%%%%%%%%%%%%%%%%%%%%%%%%%%%%%%%%%%%%%%%%%%%%%%%%%%%%%%%%%%%%%%%%%%%%%%%%
%%%%%%%%%%%%%%%%%%%%%%%%%%%%%%%%%%%%%%%%%%%%%%%%%%%%%%%%%%%%%%%%%%%%%%%%%%
\section{Conclusion}
Important progress was achieved recently in the field of radio detection of high energy cosmic rays. A complete understanding of the electric field production mechanism is necessary to go back from the radio signals to the primary cosmic ray characteristics (direction, energy, nature). Different theoretic approaches of the radio emission are under investigation, some of which predict at first order a linear dependence of the radio electric field with the vector cross product $-\mathbf{v}\times\mathbf{B}$.

This dependence offers a simple interpretation of the experimental observations of the CODALEMA experiment in two polarizations, such as the anisotropy of detection efficiency close to the threshold or the signs of the signals in each polarization. This polarization factor is a major effect to take into account when observing cosmic rays with radio antennas.
%%%%%%%%%%%%%%%%%%%%%%%%%%%%%%%%%%%%%%%%%%%%%%%%%%%%%%%%%%%%%%%%%%%%%%%%%%%%%%%%%%%%%%%

\end{document}